\documentclass[10pt,american,english]{IEEEtran}
\usepackage[T1]{fontenc}
\usepackage[latin9]{inputenc}
\usepackage{amsthm}
\usepackage{graphicx}

\makeatletter
  \theoremstyle{plain}
  \newtheorem{thm}{Theorem}
 \ifx\proof\undefined\
   \newenvironment{proof}[1][\proofname]{\par
     \normalfont\topsep6\p@\@plus6\p@\relax
     \trivlist
     \itemindent\parindent
     \item[\hskip\labelsep
           \scshape
       #1]\ignorespaces
   }{%
     \endtrivlist\@endpefalse
   }
   \providecommand{\proofname}{Proof}
 \fi

\@ifundefined{showcaptionsetup}{}{%
 \PassOptionsToPackage{caption=false}{subfig}}
\usepackage{subfig}
\makeatother

\usepackage{babel}

\begin{document}

\title{A Strictly Improved Achievable Region for Multiple Descriptions Using
Combinatorial Message Sharing}

\author{Kumar Viswanatha, Emrah Akyol and Kenneth Rose\\
ECE Department, University of California - Santa Barbara\\
\{kumar,eakyol,rose\}@ece.ucsb.edu\thanks{The work was supported by the NSF under grants CCF-0728986 and CCF - 1016861} }

\maketitle
\thispagestyle{empty}
\begin{abstract}
We recently proposed a new coding scheme for the L-channel multiple
descriptions (MD) problem for general sources and distortion measures
involving `Combinatorial Message Sharing' (CMS) \cite{MD_ISIT} leading
to a new achievable rate-distortion region. Our objective in this
paper is to establish that this coding scheme strictly subsumes the
most popular region for this problem due to Venkataramani, Kramer
and Goyal (VKG) \cite{VKG}. In particular, we show that for a binary
symmetric source under Hamming distortion measure, the CMS scheme
provides a strictly larger region for all L$>2$. The principle of
the CMS coding scheme is to include a common message in every subset
of the descriptions, unlike the VKG scheme which sends a single common
message in all the descriptions. In essence, we show that allowing
for a common codeword in every subset of descriptions provides better
freedom in coordinating the messages which can be exploited constructively
to achieve points outside the VKG region. \end{abstract}

\section{Introduction\label{sec:Introduction}}

The Multiple Descriptions (MD) problem has been studied extensively
since late 1970s, see \cite{EGC,ZB,VKG,Ozarow,wang,Ramchandran} and
the references therein. In a $L-$descriptions MD setup, the encoder
sends $L$ packets (descriptions) which are sent to the receiver over
$L$ different channels. In the most general setting, it is assumed
that the decoder receives a subset of the descriptions without any
error and the remaining are completely lost. The decoder reconstructs
the source upto a given level of distortion when a subset of the descriptions
are received. The goal of the MD problem is to establish the complete
rate-distortion region to trade-off the encoding rates to the achievable
distortions. The general setup has remained challenging and unsolved
due to the intricacies of the problem in maintaining the balance between
the full reconstruction quality versus quality of individual descriptions.

Until recently, for general sources and distortion measures, the most
recognized achievable rate-distortion region for the $L-$channel
MD setup was due to Venkataramani, Kramer and Goyal (VKG) \cite{VKG},
whose encoding scheme builds on the prior work for the 2-channel case
by El-Gamal and Cover (EC) \cite{EGC} and Zhang and Berger (ZB) \cite{ZB}.
The VKG scheme involves a combinatorial number of refinement codebooks
along with a single shared codebook used to control the redundancy
across the descriptions. We introduced a new encoding scheme in \cite{MD_ISIT}%
\footnote{For the benefit of the reviewers, the submitted version of \cite{MD_ISIT}
is available at : http://www.scl.ece.ucsb.edu/Kumar/ISIT\_MD\_Sub.pdf%
} involving `Combinatorial Message Sharing' (CMS) which differs from
the VKG scheme primarily in the number of shared codebooks. The CMS
scheme allows for every subset of the descriptions to share a different
common codebook, thereby leading to a combinatorial number of shared
messages. At the time of submission of \cite{MD_ISIT}, it was not
known whether the CMS scheme leads to a strictly improved rate-distortion
region over the VKG scheme. In this paper, our objective is to prove
by example that the new region is indeed strictly better. Specifically,
we show that for a binary symmetric source under Hamming distortion
measure, the CMS scheme achieves points outside the VKG region $\forall L>2$.
In fact, more generally, our result holds $\forall L>2$ for any source
and distortion measures for which the ZB scheme achives points outside
the EC scheme for the corresponding 2-descriptions problem. We note
in passing that, other encoding schemes have been proposed in the
literature for certain special cases (specific sources and distortion
measures) of the $L-$channel MD setup \cite{Ramchandran}, which
achieve points outside $\mathcal{RD}_{VKG}$. However, none of these
schemes have been proven to subsume or outperform $\mathcal{RD}_{VKG}$
for general sources and distortion measures. The potential implications
of our results on these coding schemes are beyond the scope of this
paper. In the following Section, we formally state the $L-$channel
MD setup and describe the prior results due to EC \cite{EGC}, ZB
\cite{ZB}, VKG \cite{VKG} and the CMS scheme \cite{MD_ISIT}. In
section \ref{sec:Proof-of-strict}, we prove the strict improvement
of the achievable region.

\section{Formal Definitions and Prior results\label{sec:Prior_results}}

We follow the notation in \cite{MD_ISIT}. A source produces $n$
iid copies, denoted by $X^{n}=\left(X^{(1)},X^{(2)}\ldots,X^{(n)}\right)$,
of a generic random variable $X$ taking values in a finite alphabet
$\mathcal{X}$. We denote $\mathcal{L}=\{1,\ldots,L\}$. There are
$L$ encoding functions, $f_{l}(\cdot)\,\, l\in\mathcal{L}$, which
map $X^{n}$ to the descriptions $J_{l}=f_{l}(X^{n})$, where $J_{l}\in\{1,\ldots B_{l}\}$
for some $B_{l}>0$. The rate of description $l$ is defined as $R_{l}=\log_{2}(B_{l})$.
Each of the descriptions are sent over a separate channel and are
either received at the decoder error free or are completely lost.
There are $2^{L}-1$ decoding functions for each possible received
combination of the descriptions $\hat{X}_{\mathcal{K}}^{n}=\left(\hat{X}_{\mathcal{K}}^{(1)},\ldots,\hat{X}_{\mathcal{K}}^{(n)}\right)=g_{\mathcal{K}}(J_{l}:l\in\mathcal{K})$,
$\forall\mathcal{K}\subseteq\mathcal{L},\mathcal{K}\neq\phi$, where
$\hat{X}_{\mathcal{K}}$ takes on values on a finite set $\hat{\mathcal{X}}_{\mathcal{K}}$,
and $\phi$ denotes the null set. When a subset $\mathcal{K}$ of
the descriptions are received at the decoder, the distortion is measured
as $D_{\mathcal{K}}=E\left[\frac{1}{N}\sum_{t=1}^{n}d_{\mathcal{K}}(X^{(t)},\hat{X}_{\mathcal{K}}^{(t)})\right]$
for some bounded distortion measures $d_{\mathcal{K}}(\cdot)$ defined
as $d_{\mathcal{K}}:\mathcal{X}\times\hat{\mathcal{X}}_{\mathcal{K}}\rightarrow\mathcal{R}$.
We say that a rate-distortion tuple $(R_{i},D_{\mathcal{K}}:i\in\mathcal{L},\mathcal{K}\subseteq\mathcal{L},\mathcal{K}\neq\phi)$
is achievable if there exit $L$ encoding functions with rates $(R_{1}\ldots,R_{L})$
and $2^{L}-1$ decoding functions yielding distortions $D_{\mathcal{K}}$.
The closure of the set of all achievable rate-distortion tuples is
defined as the `\textit{$L$-channel multiple descriptions RD region}'.
Note that, this region has $L+2^{L}-1$ dimensions.

In what follows, $2^{\mathcal{S}}$ denotes the set of all subsets
(power set) of any set $\mathcal{S}$ and $|\mathcal{S}|$ denotes
the set cardinality. Note that $|2^{\mathcal{S}}|=2^{|\mathcal{S}|}$.
$\mathcal{S}^{c}$ denotes the set complement. For two sets $\mathcal{S}_{1}$
and $\mathcal{S}_{2}$, we denote the set difference by $\mathcal{S}_{1}-\mathcal{S}_{2}=\{\mathcal{K}:\mathcal{K}\in\mathcal{S}_{1},\mathcal{K}\notin\mathcal{S}_{2}\}$.
We use the shorthand $\{U\}_{\mathcal{S}}$ for $\{U_{\mathcal{K}}:\mathcal{K}\in\mathcal{S}\}$%
\footnote{Note the difference between $\{U\}_{\mathcal{S}}$ and $U_{\mathcal{S}}$.
$\{U\}_{\mathcal{S}}$ is a set of variables, whereas $U_{\mathcal{S}}$
is a single variable. %
}.

\subsection{VKG Encoding Scheme}

\begin{figure}
\centering\includegraphics[scale=0.4]{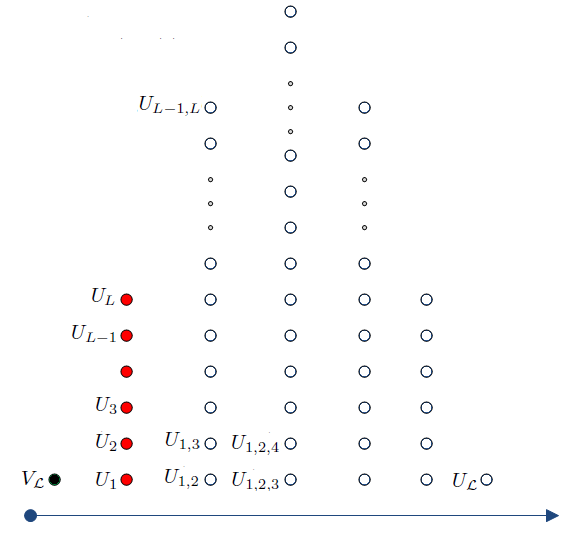}\caption{Codebook generation for VKG coding scheme. Black ($V_{\mathcal{L}}$)
indicates `common random variable'. Red ($U_{l}\,\, l\in\mathcal{L}$)
indicates `base layer random variables' and White ($U_{\mathcal{S}}$
$|\mathcal{S}|>1$) indicates `refinement random variables'. The arrow
indicates the order of codebook generation.\label{fig:VKG}}

\end{figure}

The achievable region of \cite{VKG} is denoted here $\mathcal{RD}_{VKG}$
and is described as follows. Let $(V_{\mathcal{L}},\{U\}_{2^{L}-\phi})$
be any set of $2^{L}$ random variables distributed jointly with $X$.
Then, an RD tuple is said to be achievable if there exist functions
$\psi_{\mathcal{S}}(\cdot)$ such that:\begin{eqnarray}
\sum_{l\in\mathcal{S}}R_{l} & \geq & |\mathcal{S}|I(X;V_{\mathcal{L}})-H(\{U\}_{2^{\mathcal{S}}-\phi}|X,V_{\mathcal{L}})\nonumber \\
 &  & +\sum_{\mathcal{K}\subseteq\mathcal{S}}H\left(U_{\mathcal{K}}|\{U\}_{2^{\mathcal{K}}-\phi-\mathcal{K}}\right)\label{eq:VKG_rate}\\
D_{\mathcal{S}} & \geq & E\left[d_{\mathcal{S}}\left(X,\psi_{\mathcal{S}}(V_{\mathcal{L}},\{U\}_{2^{\mathcal{S}}-\phi})\right)\right]\label{eq:VKG}\end{eqnarray}
$\forall\mathcal{S}\subseteq\mathcal{L}$. The closure of the achievable
tuples over all such $2^{L}$ random variables gives $\mathcal{RD}_{VKG}$.
Here, we only present an overview of the encoding scheme. The order
of codebook generation of the auxiliary random variables is shown
in Figure \ref{fig:VKG}. First, $2^{nR_{\mathcal{L}}^{''}}$ codewords
of $V_{\mathcal{L}}$ are generated using the marginal distribution
of $V_{\mathcal{L}}$. Conditioned on each codeword of $V_{\mathcal{L}}$,
$2^{nR_{l}^{'}}$ codewords of $U_{l}$ are generated according to
their respective conditional densities. Next, for each $j\in(1,\ldots,2^{n(R_{\mathcal{L}}+\sum_{l\in\mathcal{K}}R_{l}^{'})})$,
a single codeword is generated for $U_{\mathcal{K}}(j)$ conditioned
on $(v_{\mathcal{L}}(j),\{u(j)\}_{2^{\mathcal{K}}-\phi-\mathcal{K}})$
$\,\,\forall\mathcal{K}\subseteq\mathcal{L},\,\,|\mathcal{K}|>1$.
Note that to generate the codebook for $U_{\mathcal{K}}$, we first
need the codebooks for all $\{U\}_{2^{\mathcal{K}}-\phi-\mathcal{K}}$
and $V_{\mathcal{L}}$. 

On observing a typical sequence $x^{n}$, the encoder tries to find
a jointly typical codeword tuple one from each codebook. Codeword
index of $U_{l}$ (at rate $R_{l}^{'}$) is sent in description $l$.
Along with the `\textit{private}' messages, each description also
carries a `\textit{shared message}' at rate $R_{\mathcal{L}}^{''}$,
which is the codeword index of $V_{\mathcal{L}}$. Hence the rate
of each description is $R_{l}=R_{l}^{'}+R_{\mathcal{L}}^{''}$. VKG
showed that, to ensure finding a set of jointly typical codewords
with the observed sequence, the rates must satisfy (\ref{eq:VKG_rate}).
It then follows from standard arguments (see for example ``typical
average lemma'' \cite{Gamal_notes}) that, if the random variables
also satisfy (\ref{eq:VKG}), then the distortion constraints are
met. Note that, $V_{\mathcal{L}}$ is the \textit{only} shared random
variable. $U_{l}:l\in\mathcal{L}$ form the base layer random variables
and all $U_{\mathcal{K}}:|\mathcal{K}|\geq2$ form the refinement
layers. Observe that the codebook generation follows the order: shared
layer $\rightarrow$ base layer $\rightarrow$ refinement layer. 

The VKG scheme for the 2-descriptions scenario involves 4 auxiliary
random variables $V_{12},U_{1},U_{2}$ and $U_{12}$. The VKG region
was originally derived as an extension of the EC \cite{EGC} and ZB
\cite{ZB} coding schemes, which were designed for the 2-descriptions
scenario. The first of the two regions was by El-Gamal and Cover and
their rate region (denoted here by $\mathcal{RD}_{EC}$) is obtained
by setting $V_{12}=\Phi$ in $\mathcal{RD}_{VKG}$, where $\Phi$
is a constant. Zhang and Berger (their region is denoted here by $\mathcal{RD}_{ZB}$)
later showed that, including the shared random variable can give strict
improvement over $\mathcal{RD}_{EC}$. Their result, while perhaps
counter-intuitive at first, clarifies the fact that, a shared message
among the descriptions helps to better coordinate the messages, thereby
providing a strictly improved RD region, even though it introduces
redundancy. We will describe their result in detail in Section \ref{sec:Proof-of-strict},
as our example builds upon theirs. However, it is known that $\mathcal{RD}_{EC}$
is complete for some special cases of the setup (see for example \cite{Ozarow,Gamal_notes}).

\subsection{CMS Encoding Scheme}

\begin{figure}
\centering\includegraphics[scale=0.38]{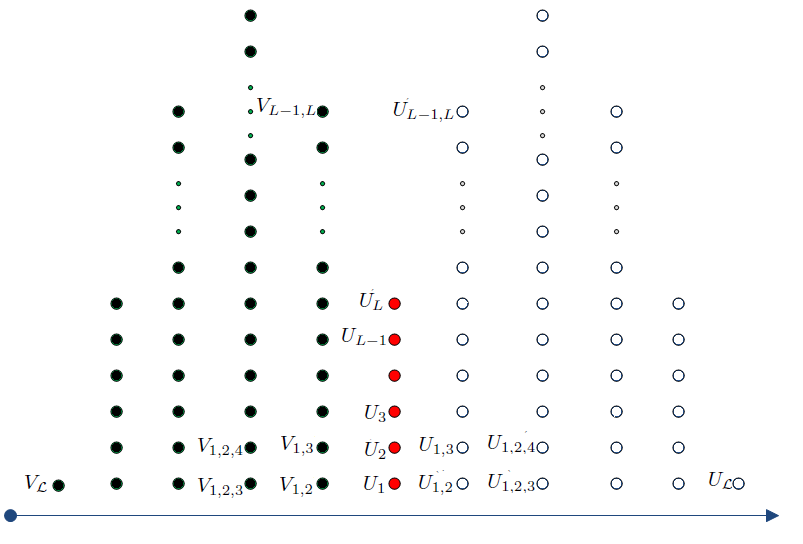}\caption{Codebook generation for the CMS coding scheme. \label{fig:L_Channel_CMS}}

\end{figure}

In this section, we briefly describe our CMS encoding scheme in \cite{MD_ISIT}.
The VKG encoding scheme employs \textit{one} common codeword ($V_{\mathcal{L}}$)
that is sent in all the $L$ descriptions. However, when dealing with
$L>2$ descriptions, restricting to a single shared message could
be suboptimal. The CMS scheme therefore allows for `combinatorial
message sharing', i.e a common codeword is sent in each (non-empty)
subset of the descriptions. Before describing the codebook generation
and stating the theorem, we define the following subsets of $2^{\mathcal{L}}$:\begin{eqnarray}
\mathcal{I}_{W} & = & \{\mathcal{S}:\mathcal{S}\in2^{\mathcal{L}},\,\,|\mathcal{S}|=W\}\nonumber \\
\mathcal{I}_{W+} & = & \{\mathcal{S}:\mathcal{S}\in2^{\mathcal{L}},\,\,|\mathcal{S}|>W\}\label{eq:Iw}\end{eqnarray}
Let $\mathcal{B}$ be any non-empty subset of $\mathcal{L}$ with
$|\mathcal{B}|\leq W$. We define the following subsets of $\mathcal{I}_{W}$
and $\mathcal{I}_{W+}$:\begin{eqnarray}
\mathcal{I}_{W}(\mathcal{B}) & = & \{\mathcal{S}:\mathcal{S}\in\mathcal{I}_{W},\,\,\mathcal{B}\subseteq\mathcal{S}\}\nonumber \\
\mathcal{I}_{W+}(\mathcal{B}) & = & \{\mathcal{S}:\mathcal{S}\in\mathcal{I}_{W+},\,\,\mathcal{B}\subseteq\mathcal{S}\}\label{eq:Iwb}\end{eqnarray}
We also define%
\footnote{We use the notation $\mathcal{A}\subseteq\mathcal{B}$ to mean $\mathcal{A}$
is subsumed in $\mathcal{B}$ and $\mathcal{A}\subset\mathcal{B}$
to mean strictly subsumed.%
}:\begin{eqnarray}
\mathcal{J}(\mathcal{K})=\bigcup_{l\in\mathcal{K}}\mathcal{I}_{1+}(l)=\{\mathcal{S}:\mathcal{S}\subseteq\mathcal{L},|\mathcal{S}|>1,|\mathcal{K}\cap\mathcal{S}|>0\}\label{eq:Jk}\end{eqnarray}

The shared random variables are denoted by `$V$'. The base and the
refinement layer random variables are denoted by `$U$'. The codebook
generation is done in an order as shown in Figure \ref{fig:L_Channel_CMS}.
First, the codebook for $V_{\mathcal{L}}$ is generated. Then, the
codebooks for $V_{\mathcal{S}}$, $|S|=W$ are generated in the order
$W=L-1,L-2\ldots2$. $2^{nR_{\mathcal{Q}}^{''}}$ codewords of $V_{\mathcal{Q}}$
are independently generated conditioned on each codeword tuple of
$\{V\}_{\mathcal{I}_{W+}(\mathcal{Q})}$. This is followed by the
generation of the base layer codebooks, i.e. $U_{l}$, $l\in\mathcal{L}$.
Conditioned on each codeword tuple of $\{V\}_{I_{1+}(l)}$, $2^{nR_{l}^{'}}$
codewords of $U_{l}$ are generated independently. Then the codebooks
for the refinement layers are formed by generating a single codeword
for $U_{\mathcal{S}},\,\,|\mathcal{S}|>1$ conditioned on every codeword
tuple of $(\{V\}_{\mathcal{J}(\mathcal{S})},\{U\}_{2^{\mathcal{S}}-\mathcal{S}})$.
Observe that the base and the refinement layers in the CMS scheme
are similar to that in the VKG scheme, except that they are now generated
conditioned on a subset of the shared codewords. 

The encoder employs joint typicality encoding, i.e., on observing
a typical sequence $x^{n}$, it tries to find a jointly typical codeword
tuple, one from each codebook. As with the VKG scheme, the codeword
index of $U_{l}$ (at rate $R_{l}^{'}$) is sent in description $l$.
However, now the codeword index of $V_{\mathcal{S}}$ (at rate $R_{\mathcal{S}}^{''}$)
is sent in \textit{all} the descriptions $l\in\mathcal{S}$. Therefore
the rate of description $l$ is:\begin{equation}
R_{l}=R_{l}^{'}+\sum_{\mathcal{K}\in\mathcal{J}(l)}R_{\mathcal{K}}^{''}\label{eq:main_rate-1}\end{equation}
We next state the main result in \cite{MD_ISIT} which describes a
new region for the $L-$Channel MD setup achievable by the CMS scheme.
Let $(\{V\}_{\mathcal{J}(\mathcal{L})},\{U\}_{2^{\mathcal{L}}-\phi})$
be any set of $2^{L+1}-L-2$ random variables jointly distributed
with $X$. We define the quantities $\alpha_{W}(\mathcal{Q})$ and
$\beta(\mathcal{S})$ as follows:

\begin{eqnarray}
 & \alpha_{W}(\mathcal{Q})= & \sum_{\mathcal{S}\in\mathcal{Q}}H\left(V_{\mathcal{S}}|\{V\}_{\mathcal{I}_{W+}(\mathcal{S})}\right)\nonumber \\
 &  & -H\left(\{V\}_{\mathcal{Q}}|\{V\}_{\mathcal{I}_{W+}},X\right)\,\,\,\forall\mathcal{Q}\subseteq\mathcal{I}_{W}\label{eq:alpha_thm}\\
 & \beta(\mathcal{S})\,\,= & \sum_{\mathcal{K}\subseteq\mathcal{S}}H\left(U_{\mathcal{K}}|\{U\}_{2^{\mathcal{K}}-\phi-\{\mathcal{K}\}},\{V\}_{\mathcal{J}(\mathcal{K})}\right)\nonumber \\
 &  & -H\left(\{U\}_{2^{\mathcal{S}}-\phi}|\{V\}_{\mathcal{I}_{1+}},X\right)\,\,\,\forall\mathcal{S}\subseteq\mathcal{L}\label{eq:beta_thm}\end{eqnarray}
We follow the convention $\alpha_{W}(\phi)=\beta(\phi)=0$. Let $R_{\mathcal{K}}^{''}$
$\forall\mathcal{K}\in\mathcal{J}(\mathcal{L})$ and $R_{l}^{'}$
$\forall l\in\mathcal{L}$ be any set of rate tuples satisfying:\begin{eqnarray*}
\sum_{\mathcal{K}\in\mathcal{Q}}R_{\mathcal{K}}^{''} & > & \alpha_{W}(\mathcal{Q})\,\,\forall\mathcal{Q}\subseteq\mathcal{I}_{W},W\in\mathcal{L}\\
\sum_{l\in\mathcal{S}}R_{l}^{'} & > & \beta(\mathcal{S})\,\,\forall\mathcal{S}\subseteq\mathcal{L}\end{eqnarray*}
then, the RD region for the $L-$channel MD problem contains the rates
and distortions for which there exist functions $\psi_{\mathcal{S}}(\cdot)$,
such that \begin{eqnarray}
R_{l} & \geq & R_{l}^{'}+\sum_{\mathcal{K}\in\mathcal{J}(l)}R_{\mathcal{K}}^{''}\label{eq:rate_condition_thm}\\
D_{\mathcal{S}} & \geq & E\left[d_{\mathcal{S}}\left(X,\psi_{\mathcal{S}}\left(\{V\}_{\mathcal{J}(\mathcal{S})},\{U\}_{2^{\mathcal{S}}-\phi}\right)\right)\right]\label{eq:dist_condition_thm}\end{eqnarray}
The closure of the achievable tuples over all such $2^{L+1}-L-2$
random variables is denoted by $\mathcal{RD}_{CMS}$. Observe that
both the VKG and the CMS schemes are same as the ZB scheme for 2 descriptions
scenario.

\section{Proof of strict improvement\label{sec:Proof-of-strict}}

Note that, the total number of auxiliary random variables in the CMS
scheme is almost twice that in the VKG scheme (which already is exponential
in $L$). At the time of submission of \cite{MD_ISIT}, it was yet
unclear if this increase pays off with an improved achievable region.
The following theorem, being the main contribution of this paper,
establishes that there exists scenarios for which $\mathcal{RD}_{CMS}$
is strictly larger than $\mathcal{RD}_{VKG}$.
\begin{thm}
(i) The rate-distortion region achievable by the CMS scheme is always
at least as large as the region achievable by the VKG region, i.e.:\begin{equation}
\mathcal{RD}_{VKG}\subseteq\mathcal{RD}_{CMS}\end{equation}
(ii) There exists scenarios for which the CMS scheme leads to a region
strictly larger than that achievable by the VKG scheme, i.e.:\begin{equation}
\mathcal{RD}_{VKG}\neq\mathcal{RD}_{CMS}\Rightarrow\mathcal{RD}_{VKG}\subset\mathcal{RD}_{CMS}\end{equation}
Specifically, for a binary symmetric source under Hamming distortion
measure, the CMS scheme achieves a strictly larger rate-distortion
region compared to the VKG scheme $\forall L>2$. \end{thm}
\begin{proof}
Part (i) of the theorem is rather simple to prove and is a straight
forward corollary of the main theorem in \cite{MD_ISIT}. It follows
directly by setting $V_{\mathcal{S}}=\Phi$ $\forall\mathcal{S}$
such that $|\mathcal{S}|<L$ in $\mathcal{RD}_{CMS}$. We then have
$\alpha_{W}(\mathcal{Q})=0$ $\forall\mathcal{Q},W<L$. Substituting
in (\ref{eq:rate_condition_thm}), we get $R_{(\mathcal{S})}\geq\beta(\mathcal{S})+|\mathcal{S}|\alpha_{L}(\mathcal{L})$
which is same as (\ref{eq:VKG}). 

We prove (ii) by considering the binary symmetric source example for
which the CMS scheme achieves points which cannot be achieved by the
VKG scheme. Note that, once we prove that the CMS scheme achieves
a strictly larger region for some $L=l>2$, then it must be true for
all $L\geq l$. Hence to prove (ii), it is sufficient for us to show
that it is true for $L=3$. However, we first include an example for
$L=4$ for building intuition and understanding of the type of scenarios
where the CMS scheme provides strict improvement. Then we will prove
the result for $L=3$. We also note that obviously scenarios exit
for which $\mathcal{RD}_{CMS}=\mathcal{RD}_{VKG}$ (for example when
$\mathcal{RD}_{VKG}$ is complete \cite{VKG}). Finding the set of
all such scenarios is an interesting problem in itself and is beyond
the scope of this paper. To describe our example, we require certain
results pertaining to binary multiple descriptions \cite{ZB} and
successive refinement of binary sources \cite{cover_SR,Rimoldi_SR}.
In what follows, we state these results.

\textbf{The Zhang-Berger example} : Zhang and Berger proved that,
for the binary symmetric 2-descriptions MD problem under Hamming distortion
measure, sending a common codeword in both the descriptions provides
a strict improvement over the EC scheme. We briefly describe their
result. Note that the rate-distortion region has 5 dimensions denoted
by $(R_{1},R_{2},D_{1},D_{2},D_{12}$). Denote the rate-distortion
region achievable by the EC scheme by $\mathcal{RD}_{EC}(bern(1/2))$
and the corresponding region achievable by the ZB scheme (i.e. achieved
by adding a common codeword among the two descriptions) by $\mathcal{RD}_{ZB}(bern(1/2)$.
Obviously $\mathcal{RD}_{EC}(bern(1/2))\subseteq\mathcal{RD}_{ZB}(bern(1/2))$,
as we can always choose not to send any common codeword in the ZB
scheme. Denote by $\bar{R}_{EC}(D)$ the following cross section of
$\mathcal{RD}_{EC}(bern(1/2))$:\begin{eqnarray}
\overline{R}_{EC}(D)=\inf\Bigl\{ R_{1}+R_{2}:\,\,\,\, D_{1}+D_{2}\leq2D\nonumber \\
(R_{1},R_{2},D_{1},D_{2},0)\in\mathcal{RD}_{EC}(bern(1/2))\Bigr\}\label{eq:EC_Crosssection}\end{eqnarray}
Similarly, denote by $\bar{R}_{ZB}(D)$, the corresponding cross section
of $\mathcal{RD}_{ZB}(bern(1/2))$. To show that $\mathcal{RD}_{EC}(bern(1/2))\subset\mathcal{RD}_{ZB}(bern(1/2))$,
they considered a particular joint probability mass function (PMF)
$P^{*}(X,V_{12},U_{1},U_{2},U_{12})$. Let us denote the achievable
region associated with this PMF by $\mathcal{RD}_{P^{*}}(bern(1/2))$
and the corresponding cross-section (\ref{eq:EC_Crosssection}) by
$\overline{R}_{P^{*}}(D)$. They showed that $\exists D^{*}>0$ such
that $\overline{R}_{EC}(D^{*})>\overline{R}_{P^{*}}(D^{*})\geq\overline{R}_{ZB}(D^{*})$.
We refer the reader to \cite{ZB} for a detailed derivation and the
values of $P^{*}$ and $D^{*}$. 

\selectlanguage{american}%
\begin{figure*}
\subfloat[\selectlanguage{english}%
\label{fig:Successive-refinement-setup}\selectlanguage{american}
]{\selectlanguage{english}%
\centering\includegraphics[scale=0.35]{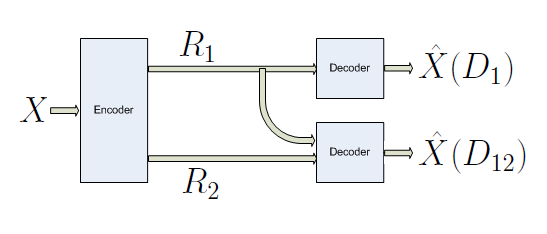}\foreignlanguage{american}{}\selectlanguage{american}
}~~~~~~~~\subfloat[\selectlanguage{english}%
\label{fig:Example-to-demonstrate}\selectlanguage{american}
]{\selectlanguage{english}%
\centering\includegraphics[scale=0.33]{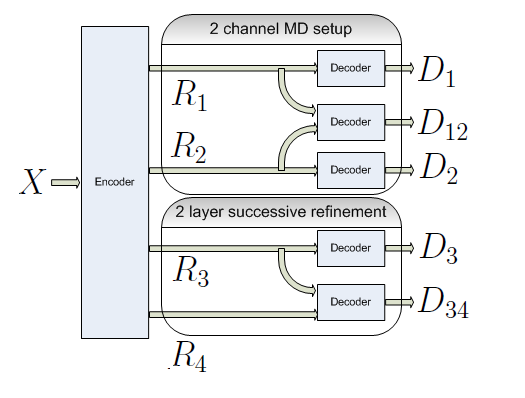}\foreignlanguage{american}{}\selectlanguage{american}
}~~~~~~~~\subfloat[\selectlanguage{english}%
\label{fig:3d_eg}\selectlanguage{american}
]{\selectlanguage{english}%
\centering\includegraphics[scale=0.33]{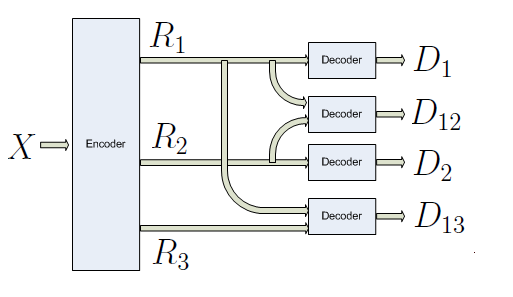}\foreignlanguage{american}{}\selectlanguage{american}
}

\caption{(a)\foreignlanguage{english}{Successive refinement setup (b) Equivalent
model for the considered cross-section of the 4-descriptions setup
(c) Equivalent model for the considered cross-section of the 3-descriptions
setup}}

\end{figure*}

\selectlanguage{english}%
\textbf{Successive refinement} : The problem of successive refinement
was first proposed by Equitz and Cover in \cite{cover_SR} and has
since then been studied extensively by information theorists \cite{cover_SR,Rimoldi_SR}.
The problem is motivated by scalable coding, where the encoder generates
two layers of information called the base layer and the enhancement
layer. The base layer provides a coarse reconstruction of the source,
while the enhancement layer is used to `refine' the reconstruction
beyond the base layer. The objective is to encode the two layers such
that the distortion at both the base and the enhancement layers are
optimal. This setup is shown schematically in Figure \ref{fig:Successive-refinement-setup}.
Observe that, the 2-layer successive refinement region is indeed a
special case (the cross-section $(R_{1}.R_{2},D_{1},\infty,D_{12})$)
of the 2-descriptions MD setup where the distortion constraint on
one of the individual descriptions is removed. The complete rate region
for successive refinement was derived in \cite{Rimoldi_SR} where
it was shown that the EC coding scheme achieves the complete rate
region. 

An interesting followup question is that of `\textit{successive refinability}'
of sources. Assume $d_{1}=d_{2}=d$, then a source is said to be successively
refinable under $d$ if $\forall D_{1}\geq D_{2}$, the rate point
$(R_{1},R_{2})=(\mathcal{RD}_{s}(D_{1}),\mathcal{RD}_{s}(D_{2})-\mathcal{RD}_{s}(D_{1}))$
is achievable, where $\mathcal{RD}_{s}(D)$ denotes Shannon's rate
distortion function. This condition implies that there is no loss
in describing the source in two successive parts. An important point
to note is that, for a successively refinable source, when the encoder
operates at $(R_{1},R_{2})=(\mathcal{RD}_{s}(D_{2}),\mathcal{RD}_{s}(D_{2})-\mathcal{RD}_{s}(D_{1}))$,
there is \textit{absolutely no redundancy} between the two layers
of information, i.e., the \textit{two layers cannot carry a common
codeword}. We finally note that a binary symmetric source is successively
refinable under Hamming measure \cite{cover_SR}.

\textbf{Proof of (ii) : $\mathbf{L=4}$} : Consider a 4-descriptions
MD problem for a binary symmetric source ($bern(\frac{1}{2})$) under
Hamming distortion measure. The rate-distortion region consists of
19 dimensions. We denote the region achievable by the VKG scheme by
$\mathcal{RD}_{VKG}^{4}$ and that achievable using the CMS scheme
by $\mathcal{RD}_{CMS}^{4}$. We now consider a particular cross-section
of these regions where we apply constraints only on $D_{1},D_{2},D_{12},D_{3}$
and $D_{34}$. We remove the constraints on all other distortions,
i.e. we set $D_{4},D_{13},D_{14},D_{23},D_{24}D_{123},D_{124},D_{134},D_{234}$
and $D_{1234}$ to $\infty$. Equivalently, we can think of a 4 descriptions
MD problem with a particular channel failure pattern, wherein only
one of the following sets of descriptions can reach the decoder reliably:
$(1,2,3,\{1,2\},\{3,4\})$ as shown in Figure \ref{fig:Example-to-demonstrate}.
We denote the set of all achievable points for this setup using the
VKG and the CMS schemes by $\mathcal{\widetilde{RD}}_{VKG}^{4}$ and
$\mathcal{\widetilde{RD}}_{CMS}^{4}$ respectively. Note that, this
equivalent model is used simply for analysis purposes, while we are
actually interested in a cross section of the general binary symmetric
4-descriptions region. 

Observe that, with respect to the first 2 descriptions, we have a
simple 2-descriptions problem and with respect to the last 2 descriptions,
we have a successive refinement problem. Extending the arguments of
Zhang and Berger, we define the following infimum of $\mathcal{\widetilde{RD}}_{VKG}^{4}$
:\begin{eqnarray}
\tilde{R}_{VKG}^{4}(D)=\inf\Bigl\{ R_{1}+R_{2}:D_{1}+D_{2}\leq2D,\nonumber \\
(R_{1},R_{2},R_{3},R_{4},D_{1},D_{2},0,D_{3},D_{34})\in\mathcal{\widetilde{RD}}_{VKG}^{4},\nonumber \\
R_{3}=\mathcal{RD}_{s}(D_{3}),\,\, R_{3}+R_{4}=\mathcal{RD}_{s}(D_{34})\Bigr\}\label{eq:example_cross_VKG-1}\end{eqnarray}
Denote the corresponding infimum of $\mathcal{\widetilde{RD}}_{CMS}^{4}$
by $\tilde{R}_{CMS}^{4}(D)$. Recall that the VKG scheme forces all
the descriptions to have a single common codeword. Constraints $R_{3}=\mathcal{RD}_{s}(D_{3})$
and $R_{3}+R_{4}=\mathcal{RD}_{s}(D_{34})$ ensure that descriptions
3 and 4 carry completely complementary information, i.e. they \textit{cannot}
carry a common codeword %
\footnote{Note that, the constraint $R_{3}=\mathcal{RD}_{s}(D_{3})$ is in fact
redundant. Just the constraint $R_{3}+R_{4}=\mathcal{RD}_{s}(D_{34})$
is sufficient to establish that descriptions 3 and 4 cannot carry
a common codeword. However, as a binary source is successively refinable,
we can always achieve $D_{3}=\mathcal{RD}_{s}^{-1}(R_{3})$ and hence
the constraint $R_{3}=\mathcal{RD}_{s}(D_{3})$ gets applied implicitly
once we apply $R_{3}+R_{4}=\mathcal{RD}_{s}(D_{34})$. This implies
that, the gains due to the CMS scheme are not only restricted to successively
refinable sources. In fact, the CMS scheme can achieve points outside
the VKG region for any source and distortion measure for which the
ZB scheme achieves points outside the EC scheme for the corresponding
2-descriptions setup. %
}. As VKG coding scheme forces the \textit{same common codeword} among
all the 4 descriptions, it follows that:\begin{equation}
\tilde{R}_{VKG}^{4}(D^{*})=\overline{R}_{EC}(D^{*})>\overline{R}_{ZB}(D^{*})\end{equation}
On the other hand, the CMS scheme allows for distinct common codewords
to be sent in each subset of the descriptions. Hence, we can send
a common codeword only among the two descriptions 1 and 2 while still
maintaining $R_{3}=\mathcal{RD}_{s}(D_{3})\mbox{ and }R_{3}+R_{4}=\mathcal{RD}_{s}(D_{34})$.
This is achieved by setting all the common random variables to $\Phi$
except $V_{12}$, which has joint PMF $P^{*}$ with $(X,U_{1},U_{2},U_{12})$.
This allows us to achieve: \begin{equation}
\tilde{R}_{CMS}^{4}(D^{*})=\overline{R}_{ZB}(D^{*})\end{equation}
This implies that $\tilde{R}_{VKG}^{4}(D^{*})>\tilde{R}_{CMS}^{4}(D^{*})$
and hence $\mathcal{RD}_{VKG}^{4}\subset\mathcal{RD}_{CMS}^{4}$.
This example clearly illustrates the freedom the CMS scheme exhibits
in controlling the redundancy across the messages. 

\textbf{$\mathbf{L=3}$} : In similar lines to the 4-descriptions
case, we next consider a 3-descriptions MD problem for a binary symmetric
source under Hamming distortion measure. Let the achievable regions
be denote by $\mathcal{RD}_{VKG}^{3}$ and $\mathcal{RD}_{CMS}^{3}$
respectively. We consider the cross-sections of the achievable regions
where we apply constraints only on $D_{1},D_{2},D_{12}$ and $D_{13}$
as shown in Figure \ref{fig:3d_eg}. We denote these cross-sections
by $\mathcal{\widetilde{RD}}_{VKG}^{3}$ and $\mathcal{\widetilde{RD}}_{CMS}^{3}$
respectively. Now consider any point $(R_{1},R_{2},R_{3},D_{1},D_{2},D_{12},D_{13})\in\mathcal{RD}_{VKG}^{3}$
such that $(R_{1},R_{2},D_{1},D_{2},D_{12})\in\mathcal{RD}_{ZB}-\mathcal{RD}_{EC}$
and $D_{13}<D_{1}$, where for two sets $\mathcal{A}$ and $\mathcal{B}$,
$\mathcal{A}-\mathcal{B}=\{l:l\in\mathcal{A},l\notin\mathcal{B}\}$.
From the results of Zhang and Berger, if $(R_{1},R_{2},D_{1},D_{2},D_{12})\in\mathcal{RD}_{ZB}-\mathcal{RD}_{EC}$,
descriptions 1 and 2 \textit{must} carry a common codeword. Let the
rate of the common codeword be $R_{c}>0$. VKG scheme forces this
codeword to be sent as part of $R_{3}$ as well. As this common codeword
is received as part of both descriptions 1 and 3, it is redundant
in $R_{3}$ to achieve $D_{13}$. This implies that $(R_{1},R_{2},R_{3}-R_{c},D_{1},D_{2},D_{12},D_{13})\in\mathcal{RD}_{CMS}^{3}$.
As there exit points in the boundary of $\mathcal{RD}_{VKG}^{3}$
which satisfy $(R_{1},R_{2},D_{1},D_{2},D_{12})\in\mathcal{RD}_{ZB}-\mathcal{RD}_{EC}$,
the CMS scheme achieves points outside the VKG scheme. Hence, we have
shown that for a binary symmetric source under Hamming distortion
measure, the CMS scheme achieves a strictly larger rate-distortion
region than the VKG scheme for all $L>2$. 
\end{proof}

\section{Conclusion}

We recently proposed a new encoding scheme for the general $L-$channel
multiple descriptions problem involving `\textit{combinatorial message
sharing}' (CMS) which leads to a new achievable region subsuming the
most well known region for this problem by Venkataramani, Kramer and
Goyal (VKG) \cite{VKG} for general sources and distortion measures.
In this paper, we showed that there exists scenarios (particularly
for a binary symmetric source under Hamming distortion measure) for
which, the new region is strictly larger than that achievable by the
VKG scheme. As part of future work, we will investigate under what
scenarios the CMS scheme achieves the complete RD region.

\end{document}